\documentclass[twocolumn]{aastex631}

\usepackage{graphics,epsf}
\usepackage[utf8]{inputenc}
\usepackage{amsmath}                % American Mathematical Society package
\usepackage{amsfonts}               % American Mathematical Society fonts
\usepackage{amssymb}                % American Mathematical Society symbol
\usepackage{epsfig}                 % EPS figures
\usepackage{graphicx}
\usepackage{float}
\usepackage{color}

\usepackage[para,online,flushleft]{threeparttable}

\newcommand{\cm}{{~\rm cm}}
\newcommand{\km}{{~\rm km}}
\newcommand{\s}{{~\rm s}}

\newcommand{\g}{{~\rm g}}
\newcommand{\G}{{~\rm G}}
\newcommand{\K}{{~\rm K}}
\newcommand{\erg}{{~\rm erg}}

\definecolor{brown1}{rgb}{0.55, 0.27, 0.08}
\definecolor{orange1}{rgb}{1, 0.4, 0}
\definecolor{Si28S32}{rgb}{0.35, 0.35, 0.35}
\definecolor{O16}{rgb}{0.4, 0.47, 0.6}
\definecolor{Ni56}{rgb}{0.55 0.58 0.49}
\definecolor{pink1}{rgb}{1 0 1}
\definecolor{golden1}{rgb}{0.9290 0.6940 0.1250}

\begin{document}

\title{Supplying angular momentum to the jittering jets explosion mechanism using inner convection layers}
\date{June 2021}

\author{Dmitry Shishkin}
\affiliation{Department of Physics, Technion, Haifa, 3200003, Israel; s.dmitry@campus.technion.ac.il; soker@physics.technion.ac.il}

\author[0000-0003-0375-8987]{Noam Soker}
\affiliation{Department of Physics, Technion, Haifa, 3200003, Israel; s.dmitry@campus.technion.ac.il; soker@physics.technion.ac.il}
\affiliation{Guangdong Technion Israel Institute of Technology, Shantou 515069, Guangdong Province, China}

\begin{abstract}
We conduct one-dimensional stellar evolution simulations in the mass range $13-20 M_{\odot}$ to late core collapse times and find that an inner vigorous convective zone with large specific angular momentum fluctuations appears at the edge of the iron core during the collapse. The compression of this zone during the collapse increases the luminosity there and the convective velocities, such that the specific angular momentum fluctuations are of the order of $j_{\rm conv} \simeq 5 \times 10^{15} \cm^2 \s^{-1}$. If we consider that three-dimensional simulations show convective velocities that are three to four times larger than what the mixing length theory gives, and that the spiral standing accretion shock instability in the post-shock region of the stalled shock at a radius of $\simeq 100 \km$ amplify perturbations, we conclude that the fluctuations that develop during core collapse are likely to lead to stochastic (intermittent) accretion disks around the newly born neutron star. In reaching this  conclusion we also make two basic assumptions with uncertainties that we discuss. Such intermittent disks can launch jets that explode the star in the frame of the jittering jets explosion mechanism.   
\end{abstract} 

% ==========================================================
\section{INTRODUCTION}
\label{sec:intro}
% ==========================================================

Both the delayed neutrino explosion mechanism \citep{BetheWilson1985} and the jittering jets explosion mechanism \citep{Soker2010} of core collapse supernovae (CCSNe) require the presence of perturbations in the collapsing core. In the neutrino delayed mechanism the perturbations serve to break the spherical symmetry as a non-spherical flow eases the revival of the stalled shock (e.g., \citealt{CouchOtt2013, OConnorCouch2018, Mulleretal2019Jittering, Couchetal2020, KazeroniAbdikamalov2020}). In the jittering jets explosion mechanism the perturbations serve as the source of stochastic angular momentum fluctuations that facilitate the formation of intermittent accretion disks that in turn launch the jittering jets (e.g., \citealt{PapishSoker2011,  GilkisSoker2014, GilkisSoker2015, Quataertetal2019}). The seeds of all these perturbations in non-rotating cores are the convective zones in the collapsing core.

The collapsing core material encounters the stalled shock at a radius of $\simeq 100 \km$. Instabilities in the post-shock region amplify the convective-triggered perturbations in the collapsing core material. In the jittering jets explosion mechanism (e.g., \citealt{Soker2019SASI, Soker2019JitSim}) the main instability is the spiral standing accretion shock instability (spiral SASI; e.g., \citealt{BlondinMezzacappa2007, Iwakamietal2014, Kurodaetal2014, Fernandez2015, Kazeronietal2017} for studies of the SASI, and, e.g.,  \citealt{Andresenetal2019, Walketal2020, Nagakuraetal2021, Shibagakietal2021},  for recent simulations that demonstrate the spiral SASI).  

The key challenge of the jittering jets explosion mechanism is to supply large enough perturbations to seed the spiral SASI such that the final angular momentum fluctuations form intermittent accretion disks (or belts, \citealt{SchreierSoker2016}). When accretion is from the helium-rich shell in the core  (e.g., \citealt{GilkisSoker2014}) or from the  hydrogen-rich envelope  (e.g., \citealt{Quataertetal2019, AntoniQuataert2021}) that are at large distances from the center, the angular momentum fluctuations are large enough by themselves to form intermittent accretion disks. In these cases the large masses inner to these shells imply the formation of a black hole remnant. We here concentrate on explosions that leave a neutron star (NS) remnant. 

Analyzing some studies of the neutrino delayed explosion mechanism (e.g., \citealt{Mulleretal2017, Mulleretal2018Bipol, Mulleretal2019Jittering}),  \cite{Soker2019JitSim} argues that there is a mutual influence between stochastic angular momentum accretion and neutrino heating that operate together to explode CCSNe by bipolar outflows that change directions, namely, jittering jets. Nonetheless, there are clear different predictions of the two explosion mechanisms (e.g., \citealt{GofmanSoker2020}).

The question then is where in the collapsing core one has to introduce the perturbations and how large these fluctuations should be. 

\citealt{GofmanSoker2020} present the core convective velocity when the collapsing speed was $v_{\rm MF} \simeq 1000 \km \s^{-1}$.
They did not follow the evolution. 
\cite{FieldsCouch2020, FieldsCouch2021} follow the evolution of the convective velocity in one dimension (1D) and in 3D during the early phase of the collapse. They do not refer to the role of angular momentum and do not follow to late core collapse phases. With the stellar evolutionary code \textsc{mesa} we follow the evolution of the convection deep into the collapse phase of stellar models with initial masses of $M_{\rm ZAMS}=13-20 M_\odot$ (section \ref{sec:Collapse}) and emphasize the role of angular momentum perturbations (section \ref{sec:Discuss}).

% ==================================
\section{Convection zones at core collapse}
\label{sec:Collapse} % 
% ==================================

\begin{figure*}
\begin{center}
\includegraphics[trim=8cm 1.4cm 8cm 1.64cm,scale=0.43]{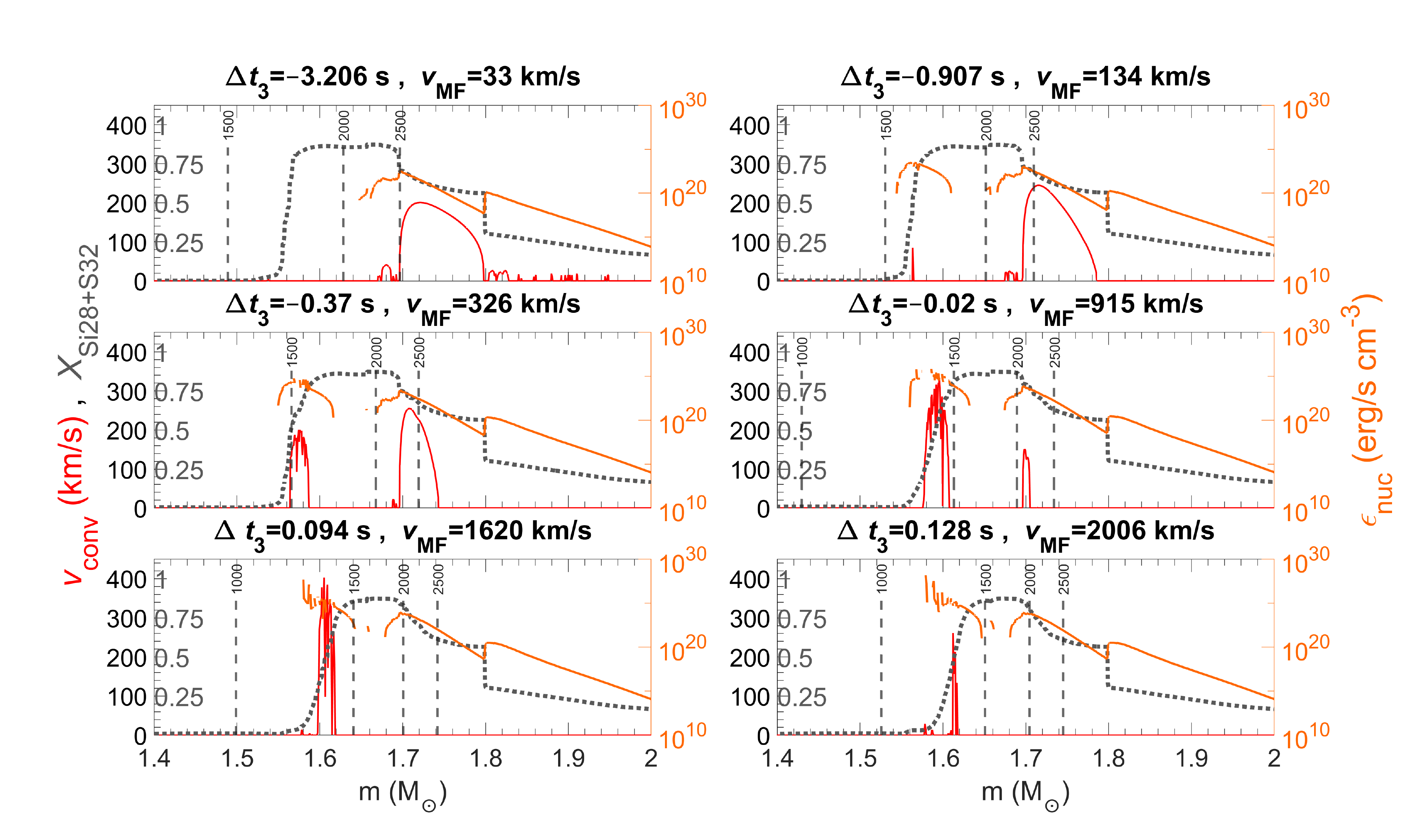}
\caption{Evolution of the mass layer $m=1.4-2 M_\odot$ of a stellar model with $M_{\rm ZAMS}=15M_\odot$ at six times from early to late core collapse. The time $\Delta t_3$ is measured relative to the time when the maximum infall velocity is $v_{\rm{MF}}=1000 \km s^{-1}$. 
We present the convective velocity (red solid line), the relative mass fraction of the $^{28}$Si and $^{32}$S combined (dotted gray line; scale on the left inside panels), and the nuclear power in $\erg \s^{-1} \cm^{-3}$ (solid orange line). 
Vertical dashed lines indicate 4 radii as we mark in km. The $r=1000 \km$ line appears only in the last 3 panels. Note the disappearance of the outer convective zone where oxygen burns and the appearance of a vigorous convective zone further in where silicon burns.}
\label{fig:6panels}
\end{center}
\end{figure*}

We simulate stellar evolution with \textsc{mesa} (version 10398; \citealt{Paxtonetal2010, Paxtonetal2013, Paxtonetal2015, Paxtonetal2018, Paxtonetal2019}).
Most of our numerical assumptions and settings are similar to those that \cite{FieldsCouch2020} used. We use the \textit{'approx21'} nuclear reaction network which tracks 21 isotopes.
We enable the Ledoux criterion (\citealt{Henyey1965}), with the mixing length parameter $\alpha_{\rm{MLT}}=1.5$ and semi-convection parameter $\alpha_{\rm{SC}}=0.01$. Overshoot follows \cite{Herwig2000} with the parameters $f=0.004$ and $f_0=0.001$.
After the iron core mass grows to $M_{\rm{core}}^{\rm{Fe}}>1.2M_{\odot}$, we limit the time step to be $\Delta t < 2 s$. We also enforce a maximum mass in each numerical shell of $\Delta m=10^{-4}M_{\rm{star}} \simeq 10^{-3} M_\odot$.
Initial metalicity is $Z=0.02$. We disable rotation.
In Fig. \ref{fig:6panels} we present some properties of the convective zones of the inner core of a $M_{\rm ZAMS}=15 M_\odot$ model at 6 times from early to late core collapse. 
We label each panel by the maximum infall velocity $v_{\rm MF}$ at the respective time, and with the time $\Delta t_3$, which is the time relative to when the maximum infall velocity is $v_{\rm MF}=1000 \km \s^{-1}$. 
In each panel we show the convective velocity according to the mixing length theory (MLT; $v_{\rm conv}$; red solid line), the relative mass fraction of the $^{28}$Si and $^{32}$S combined ($X_{28/32}$; dotted gray line), and the nuclear power per unit volume ($\epsilon_{\rm nuc}$; solid orange line). 
% FFFFFFFFFFFFFFFFFFFFFFFFFFFFFFFFFFFFFFFFFFFFFFFFFFFFFFFFFFFFF
%  trim={<left> <lower> <right> <upper>}
% FFFFFFFFFFFFFFFFFFFFFFFFFFFFFFFFFFFFFFFFFFFFFFFFFFFFFFFFFFFFF

In the first panel of Fig. \ref{fig:6panels} there is a wide vigorous convection zone in the oxygen burning layer ($m \simeq 1.7-1.8 M_\odot$ at that particular time). At later times, as the consecutive panels show, as the core accelerates its collapse a vigorous convection zone appears further in where Si/S burn (steep change in the value of  $X_{28/32}$). We see there also the large increase in the nuclear power $\epsilon_{\rm nuc}$. 
At the last panel the convection zones weaken and even completely disappear. The degree of weakening convection and the exact time of disappearance depends on the numerical resolution. In other words, we cannot handle this very late collapsing phase accurately. In any case, as we discuss in section \ref{sec:Discuss} we expect the perturbations to survive until the collapsing material encounters the stalled shock at around $r\simeq 100 \km$. There, the SASI will increase the relative perturbations further (section \ref{sec:intro}). 

\begin{figure*}
\begin{center}
 \includegraphics[trim=8cm 1.4cm 8cm 1.65cm,scale=0.43]{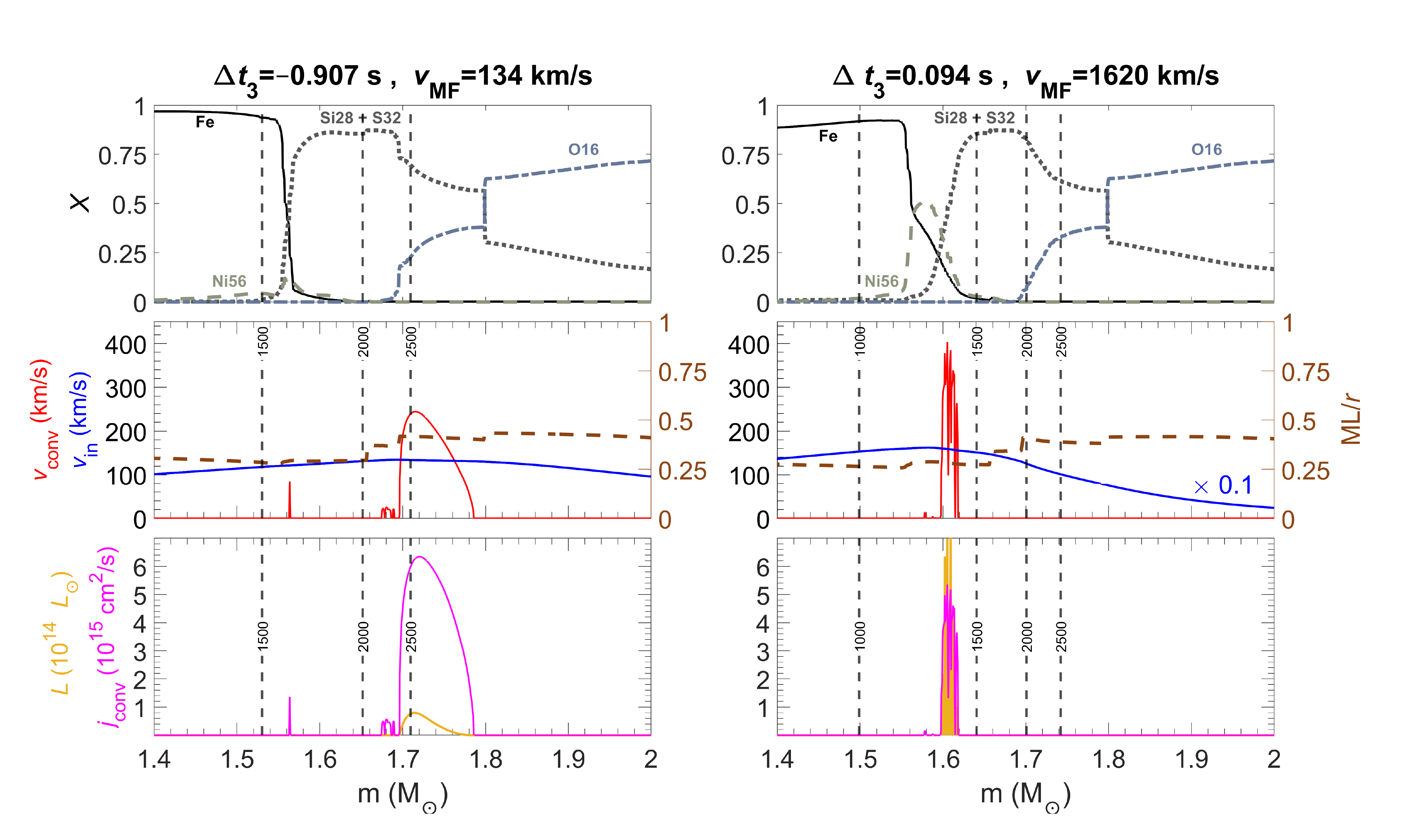}%[trim=8cm 1cm 8cm 1.56cm,scale=0.47]
\caption{Some more details of the mass layer as in figure \ref{fig:6panels} at two times as indicated on top. 
The upper panels present the mass fraction of four isotope groups as indicated. The middle panels present the convective velocity (solid red line), the ratio of mixing length to radius (dashed brown line; scale on right), and the collapsing velocity (solid blue line; note that in the middle right panel the values are of $0.1v_{\rm in}$).
The lower panels present the luminosity (yellow line; note the units of $10^{14} L_\odot$), and the specific angular momentum fluctuations due to convection $j_{\rm conv} = v_{\rm conv} r$.
Vertical lines as in Fig. \ref{fig:6panels}. 
Note the large values of the specific angular momentum perturbations $j_{\rm conv}$. }
\label{fig:DetailedFig}                                                            
\end{center}
\end{figure*}

In Fig. \ref{fig:DetailedFig} we present some more properties at $\Delta t_3=-0.907 \s$ (left column) and at $\Delta t_3=+0.094 \s$ (right column). In the upper panels we present the mass fraction of four isotope groups as we indicate inside the panels. These serve to clearly relate the convective zones to nuclear burning. 
In the middle panels we present the convective velocity (solid red line, as in Fig. \ref{fig:6panels}), the collapse velocity (inward) $v_{\rm in}$ (solid blue line), and the mixing length divided by the radius (ML/$r$; brown dashed line). From the middle panels we learn that the inner convective zone more or less coincides with the maximum inflow velocity. Below we show that this causes a large density increase. The large ratio of mixing length to radius implies large convective cells, as indeed 3D simulations find (e.g., \citealt{GilkisSoker2016, FieldsCouch2021}). Large convective cells prevent smearing the angular momentum fluctuations (section \ref{sec:Discuss}).
In the lower panel we present the luminosity (solid yellow line) and the quantity $j_{\rm conv}=v_{\rm conv} r$ (solid magenta line) which represents the specific angular momentum fluctuations. The specific angular momentum fluctuations play the key role in the jittering jets explosion mechanism. We learn also from the lower panels that vigorous convection occurs where the luminosity is large.  
% FFFFFFFFFFFFFFFFFFFFFFFFFFFFFFFFFFFFFFFFFFFFFFFFFFFFFFFFFFFFF
%  trim={<left> <lower> <right> <upper>}
% FFFFFFFFFFFFFFFFFFFFFFFFFFFFFFFFFFFFFFFFFFFFFFFFFFFFFFFFFFFFF

\begin{figure}
\begin{center}
 \includegraphics[trim=9cm 2cm 8cm 2.2cm,scale=0.5]{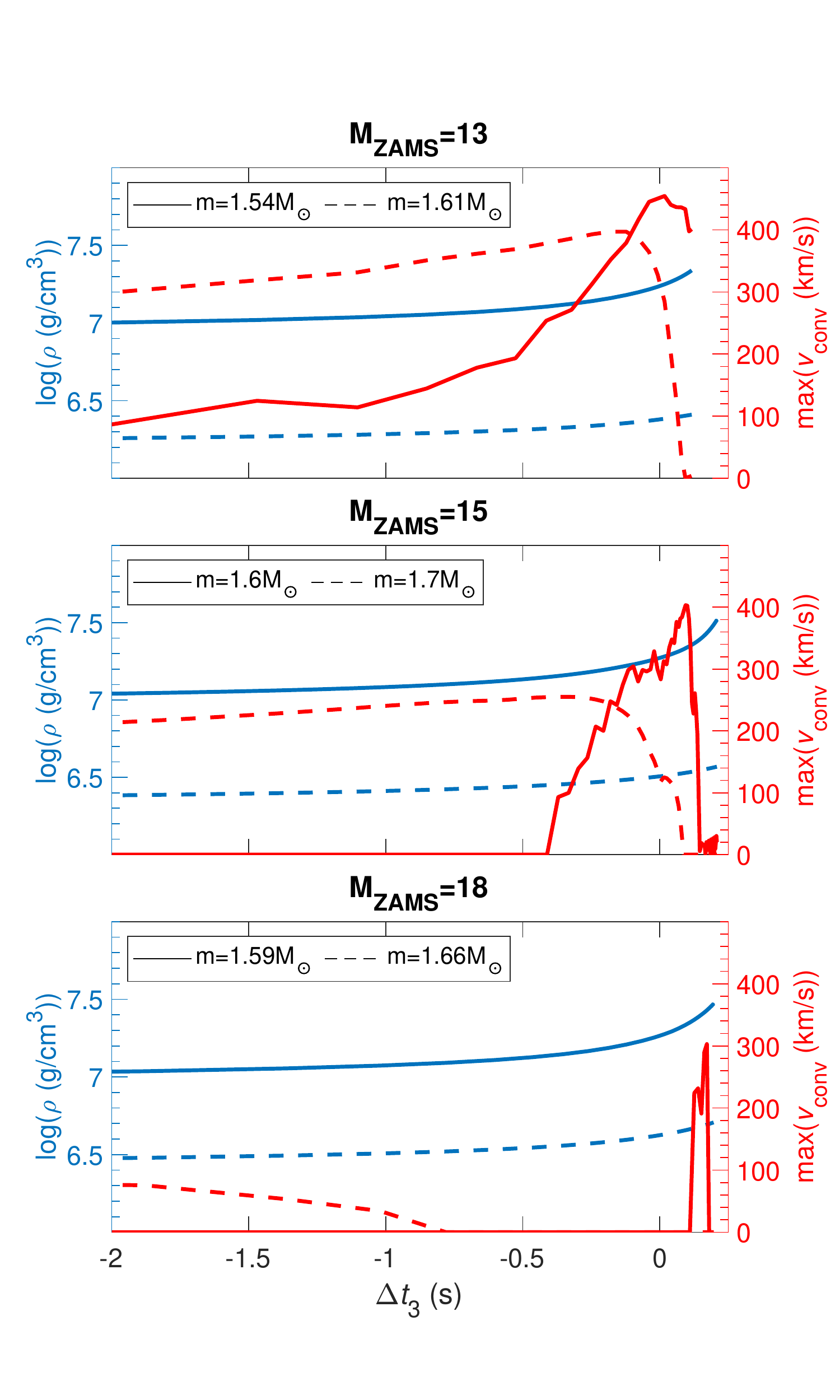} %[trim=9cm 1.5cm 8cm 1.6cm,scale=0.65] [trim=9cm 1.5cm 8cm 1.6cm,scale=0.55]
\caption{The evolution with time of density (blue lines) and maximum convective velocity (red lines) near two mass coordinates in the outer convective zone (dashed lines) and in the inner convective zone (solid lines) that we study, and for three stellar models. The densities are exactly at the mass coordinates that we give in the insets. The convective velocities are the maximum convective velocities in the respective mass coordinate vicinity ($\pm 0.05 M_\odot$). These panels emphasize the appearance of the inner convective zone as the mass there is substantially compressed during collapse.}
\label{fig:MassesFig}                                                                                     
\end{center}
\end{figure}

We find similar behavior, namely, the appearance of an inner convective zone during the collapse and the disappearance of the main early convective zone where oxygen burns, in the mass range $13-20 M_\odot$. We will explore the behavior of lower and higher mass stars in forthcoming studies. We present here the behavior of two stellar models in addition to $M_{\rm ZAMS}=15M_\odot$. In Fig. \ref{fig:MassesFig} we present the evolution with time of the convective velocity and of the density in the regions of the early convective zone where oxygen burns, and in the inner convective zone that appears later. The exact mass coordinate changes a little with time. In the inset we give the approximate mass coordinate of the two zones for the three stellar models.    

% FFFFFFFFFFFFFFFFFFFFFFFFFFFFFFFFFFFFFFFFFFFFFFFFFFFFFFFFFFFFF
%  trim={<left> <lower> <right> <upper>}
% FFFFFFFFFFFFFFFFFFFFFFFFFFFFFFFFFFFFFFFFFFFFFFFFFFFFFFFFFFFFF

Figure \ref{fig:MassesFig} emphasize the disappearance of the main convective zone where oxygen burns (dashed red line), and the appearance of the vigorous convection in the inner zone (solid red line). The figure also demonstrates the larger compression of the zone where the inner convection zone appears (solid blue line) relative to the outer convective zone. 

% ==================================
\section{Discussion and summary}
\label{sec:Discuss} % 
% ==================================

Our \textsc{mesa} simulations show that for the initial stellar mass range of ${M_{\rm ZAMS} \simeq 13-20 M_{\odot}}$ an inner vigorous convective zone appears at the beginning of collapse (when $v_{MF}>100 \km \s^{-1}$) reaching velocities of $v_{\rm conv} \simeq 300-400 \km \s^{-1}$. It corresponds to silicon (and sulfur) burning on the edge of the iron core.
The collapsing core compresses this zone (Fig. \ref{fig:MassesFig}), increasing the luminosity there (lower panels of Fig. \ref{fig:DetailedFig}), and therefore the convective velocity (Figs. \ref{fig:6panels} and \ref{fig:DetailedFig}). 
This by itself is not a new result. However, we present the specific angular momentum fluctuations $j_{\rm conv}$ and followed the evolution to late collapse time. 
The importance of this inner convective zone is that it has large values of $j_{\rm conv} \simeq 5 \times 10^{15} \cm^2 \s^{-1}$ (lower right panel of Fig. \ref{fig:DetailedFig}).

Our results are for the 1D models that we simulated with \textsc{MESA}. If we incorporate into our findings the recent results of \cite{FieldsCouch2021} who find from their 3D simulations that the angle-average convective speeds near collapse are three to four times larger than the values that \textsc{MESA} gives, we conclude that the specific angular momentum fluctuations are $j_{\rm 3d} \simeq (3-4)j_{\rm conv} \simeq 1.5-2 \times 10^{16} \cm^2 \s^{-1}$. As well, they find that low spherical harmonic indices $l=1-3$ are most powerful, i.e., large convective cells. The 1D results are compatible with this as the mixing length is ${\rm ML} \simeq 0.3r - 0.4 r$ (middle panels of Fig. \ref{fig:DetailedFig}).
The flow structure with large convective cells works for the jittering jets explosion mechanism as only few convective cells supply the mass at every dynamical time of the newly born NS $\Delta t_{\rm d} \simeq 0.01 \s$, which is the typical time for stochastic variations. Namely, the large cells prevent a smearing of the angular momentum fluctuations due to the addition of many convective cells. We emphasize that we differ from \cite{GilkisSoker2015} and from \cite{Quataertetal2019} in that we do not average over spherical shells in the convective zones. Such averaging gives a value for the random angular momentum that is about an order of magnitude lower than what we quote here (e.g., eq. 2 of \citealt{Quataertetal2019}). The new version of the jittering jets explosion mechanism (e.g., \citealt{Soker2019SASI}) assumes that the spiral-SASI modes amplify the local perturbations because the spiral-SASI largely deviates from spherical symmetry.
Therefore, in considering the perturbation seeds for the spiral-SASI we should consider the local values of the angular momentum. For the final accretion onto the newly born NS we should average over some convective cells, but not over an entire spherical shell. This reduces the angular momentum but by less than an order of magnitude. We take it that the spiral-SASI more than compensates for this reduction by averaging, something that future high-resolution 3D simulations including magnetic fields will have to examine.

The minimum specific angular momentum to have a stable orbit around a NS of mass $M_{\rm NS}=1.4 M_\odot$ is $j_{\rm min}=2.2 \times 10^{16} \cm^2 \s^{-1}$, where the orbital radius is $12 \km$.  Our results for the mass range $M_{\rm ZAMS} \simeq 13-20 M_\odot$ (lower and higher masses are the subjects of forthcoming papers) are that $j_{\rm 3d} \simeq 0.5-0.9 j_{\rm min}$. 
For the jittering jets explosion mechanism there are two factors that come to play here, which are actually assumptions of the mechanism. ($i$) The spiral-SASI (section \ref{sec:intro} and paragraph above) increases the angular momentum fluctuations by a substantial factor. ($ii$) For the purpose of jets' launching, it is possible that even an accretion belt, where the angular momentum is somewhat lower than that required for a thin accretion disk, is sufficient to launch jets (e.g., \citealt{SchreierSoker2016}).
As stated, this is actually another assumption of the model, which also requires future confirmation.

Our results significantly strengthen the jittering jets explosion mechanism as they show large angular momentum perturbations to seed the spiral SASI that in turn can lead to stochastic accretion disk (belt) formation. It is this intermittent disk that launches the jittering jets.  
We encourage future numerical simulations of CCSNe to introduce velocity perturbations in an extended region in the silicon shell, including both the early convective zone in the outer silicon shell as the perturbation are likely to survive to collapse, and from the inner silicon shell where new convective zone might appear during the collapse. Such perturbations, we expect, will lead to accretion disk/belt formation, that with the more complicated inclusion of magnetic fields will lead to stochastic jets' launching as the jittering jets explosion mechanism requires. 

% ===================================================
\section*{Acknowledgments}
% ===================================================
We thank an anonymous referee for useful comments.
This research was supported by a grant from the Israel Science Foundation (769/20).

%%%%%%%%%%%%%%%%%%%%%%%%%%%
\textbf{Data availability}

The data underlying this article will be shared on reasonable request to the corresponding author. 
%%%%%%%%%%%%%%%%%%%%%%%%%%%

%\bibliography{BIB}

\end{document}